# Phase and Power Control in the RF Magnetron Power Stations of Superconducting Accelerators *

G. Kazakevich[#], R. Johnson, Muons, Inc., Batavia, IL, USA,
V. Lebedev, V. Yakovlev, Fermilab, Batavia, IL, USA,
V. Pavlov, Budker Institute of Nuclear Physics, Novosibirsk, Russia.

The capabilities of phase and power control in injection-locked magnetrons were studied with CW 2.45 GHz, 1 kW microwave oven magnetrons. This study was aimed at the investigation of the possible application of magnetrons for powering Superconducting RF (SRF) cavities in intensity-frontier accelerators. The study demonstrated that the magnetron RF phase can be controlled in a frequency band of a MHz range with low noise and precise carrier frequency stability by injection of a resonant phase-modulated signal at a power level as low as about -12 dB. Studies of control in magnetrons driven by a resonant (injection-locking) signal demonstrated wideband vector power management in a two-channel magnetron transmitter with power combining by a 3-dB hybrid and in a single-channel magnetron transmitter utilizing control of the depth of the phase modulation. Both vector methods of power control allow simultaneously the wideband phase control necessary for the powering of SRF cavities, however, using the vector methods significantly decreases the efficiency of RF source. This paper discusses a recently developed technique of power control up to 10 dB by a single-channel transmitter at the highest efficiency with the control bandwidth in the kHz range. A simple kinetic model based on the charge drift approximation describing the interaction of the drifting charge with the synchronous wave excited in magnetrons and considering the impact of the RF resonant signal injected into the magnetron on the operation of the injection-locked tube is presented and substantiated by experimental results.

## Introduction

The RF sources feeding SRF superconducting cavities in modern accelerators require wideband phase and power control to compensate for parasitic modulations inherent in superconducting cavities operation. Depending on the SRF cavity type and mode of operation its fundamental mode bandwidth may be of the same order as the bandwidth of mechanical oscillations of the cavity walls caused by microphonics, Lorentz Force Detuning (LFD) and other noises, [1, 2]. This causes large amplitude and phase deviations of the accelerating field, which may vary from cavity to cavity. Compensation of the parasitic amplitude deviation requires an individual variation of the RF power feeding each SRF cavity that can be as high as several times. Traditional RF amplifiers such as klystrons, IOTs or solid-state amplifiers are used as high-power sources. They provide power levels up to hundreds of kW or more in CW mode at a carrier frequency in the GHz range with a bandwidth in the MHz range allowing compensation of the modulations. However, the capital cost for a unit of power for the traditional RF sources is quite high, at least a few dollars per Watt, [3]. Therefore when utilizing traditional RF sources for large-scale accelerator facilities that shall deliver beam power levels of several MWs - such as next generation neutron sources or Accelerator Driven Systems (ADS) for sub-critical reactors, the capital cost of the RF system is a significant fraction of the overall accelerator project cost. In contrast, the cost of a unit of power of a commercial, L-band, CW, high-power magnetron RF source is several times less [ibid.]. Since the magnetron has a higher efficiency compared to traditional RF sources, magnetron transmitters will allow a significant reduction of both the capital and operating costs in large-scale accelerator projects.

Phase-locked magnetrons were suggested already in the past to power linacs [4]. This principally may work for normal-conducting accelerators where instability of the accelerating voltage is determined generally by instability of magnetrons. However, in superconducting accelerators the above mentioned parasitic modulations are not caused by instability of RF sources; they are inherent in SRF cavities operation. Thus, powering SRF cavities requires locking of the phase and amplitude of the accelerating field [5], while the RF source has to be managed in phase and power with a respective bandwidth of the control.

The method of phase control in magnetrons is realized by a wideband phase modulation of the driving resonant (injection-locking) RF signal [6]. Presently three methods have been suggested and tested in order to change the power of the magnetron based transmitter in accordance with rate required for powering SRF cavities. The first method is based on a vector summation of signals of two independently phase-controlled magnetrons. Their output signals are combined by a 3-dB hybrid and the power control is achieved by controlling the phase difference of the signals locking each magnetron [6]. The second vector method uses an additional control (modulation) of the phase modulation depth in a single magnetron. Such a modulation results in the RF power being distributed between the fundamental frequency and the sidebands. If the frequency of the depth modulation is much larger than the accelerating field bandwidth, the power concentrated in the sidebands is reflected from the cavity towards dummy load. Thus, changes of the modulation depth result in a power change at the fundamental frequency [7]. Since the RF sources intended for powering SRF cavities for both vector methods operate at nominal power and a part of the power is continuously redistributed into a dummy load for absorption, both vector methods provide an average relative efficiency of about 50%-70% dependent on the range of power control (a few dB) as required for SRF cavities. A novel method proposed and studied recently, [8], provides a significantly higher average relative efficiency (more than 80%) via a single magnetron with a range of power control up to 10 dB and wideband phase control. The power control in this technique is realized by variation of the magnetron current over an extended range, where the magnetron, driven by a sufficient (in magnitude) injection-locking signal, may operate at a voltage less than the threshold of self-excitation. This provides the extended range of the power control as mentioned above. The bandwidth of the power control in this technique is determined by the bandwidth of the current feedback loop in the magnetron High Voltage (HV) power supply operating as a current source. Presently the bandwidth may be up to 10 kHz without compromising the efficiency of the power supply. Features of the injection-locked magnetrons utilizing the listed methods of phase and power control were studied in experiments with 2.45 GHz, CW, 1 kW magnetrons, refs. [6-8], and compared with a simplified analytical kinetic model of the charge drift approximation. Results of the experiments in comparison to the analytical considerations are discussed here.

* Work was supported by Fermi Research Alliance, LLC under Contract No. De-AC02- 07CH11359 with the US DOE, Office of Science and collaboration Muons, Inc. - Fermilab
[#]e-mail: gkazakevitch@yahoo.com

## A wideband phase control in injection-locked magnetrons

The wideband phase control required for an SRF cavity powered by a magnetron RF source was first studied in detail using the phase modulation technique with various configurations of transmitters in the pulsed regime with CW, 2.45 GHz magnetrons [6]. A pulsed modulator could power the two magnetrons; it used a partial discharge of a storage capacitor providing pulse duration of 5 ms, with voltage droop of about 0.4% at a negligible low ripple [6]. Experiments were performed with single and 2-cascade magnetrons injection-locked by the phase-modulated signal in the setup schematically shown in Fig. 1, [ibid.]. Two magnetrons with a frequency offset of ≈4.7 MHz at a power of about 0.5 kW have been used in experiments, Each one was installed in a module presented in Fig. 2.

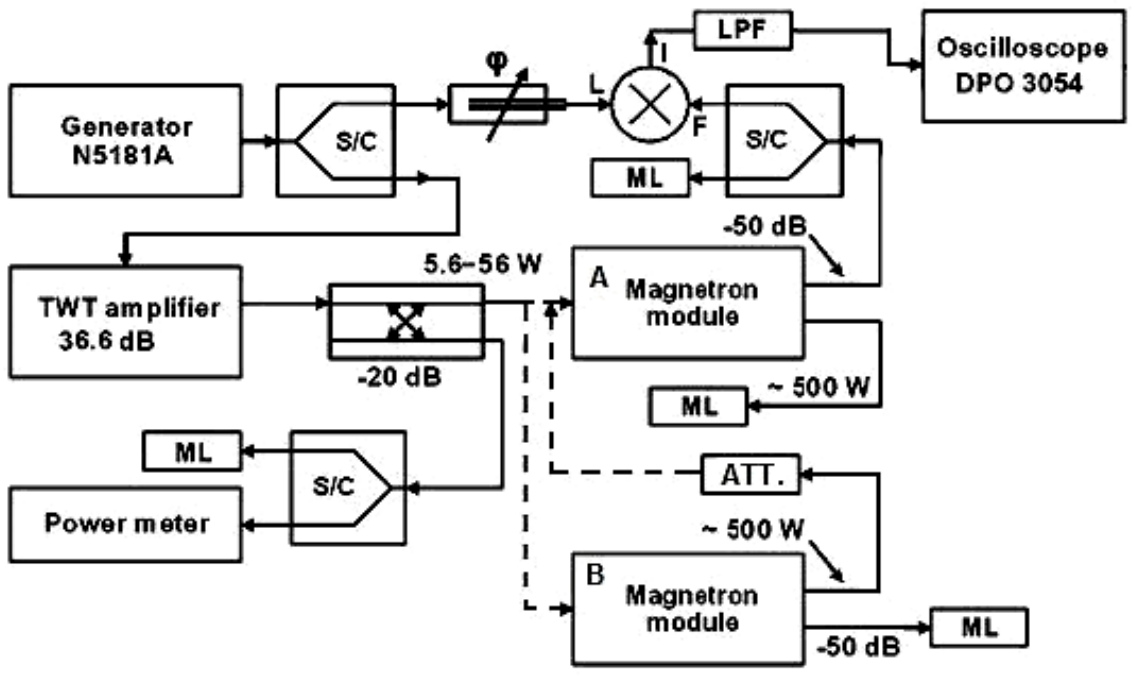

Fig. 1. Schematics for test of single and 2-cascade magnetrons frequency-locked by a phase-modulated signal. S/C is a splitter/combiner, LPF is a low pass filter, ML is a matched load, φ is a phase shifter and ATT is an attenuator.

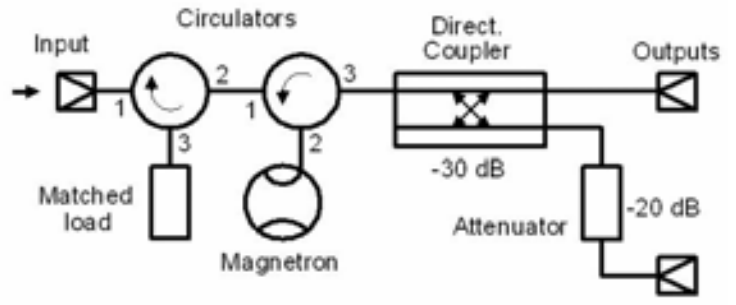

Fig. 2. Schematic of the magnetron module.

The single injection-locked magnetrons, Fig. 1, were tested in a configuration using module A, with the magnetron locked by the CW TWT amplifier and fed by the modulator, while module B was disconnected from the amplifier and the modulator. The 2-cascade magnetron was tested in a configuration in which the magnetron in module B was injection-locked by the TWT amplifier while the magnetron in module A was connected via an attenuator to the module B output. In this case the magnetron in module A was injection-locked by the pulsed signal of magnetron B, lowered in the attenuator. Both magnetrons were fed by the same modulator. Experiments demonstrated operation of the 2-cascade magnetron injection-locked at the average of the offset frequencies at attenuator values in the range of 9-20 dB [6].

Study of a wideband phase control of the single and 2-cascade injection-locked magnetrons was performed uzing internal phase modulation by a harmonic signal in the N5181A generator. The broadband (2-4 GHz) TWT amplifier did not distort the phase-modulated signal, injection-locking the magnetrons.

The transfer function magnitude characteristics (magnitude Bode plots expressing the magnitude response of the device in dependence on the frequency of the phase modulation) averaged over 8 pulses of the single and 2-cascade magnetrons, Fig. 3, were measured in the phase modulation domain by the Agilent MXAN9020A Signal Analyzer at various power of the injection-locking signal.

The measurements have been performed at the magnitude of the phase modulation of 0.07 rad., and at the magnetrons output power of $P_{Out}$ ≈450 W [6]. The measurements with the 2-cascade magnetron were carried out at the attenuator value of ≈13 dB, *i.e.*, the second cascade (tube) operated at $P_{Lock}$ ≈25 W. In Figs. 3 and 4 in the plots concerning the 2-cascade magnetron, $P_{Lock}$ denotes the power of the signal locking the first cascade.

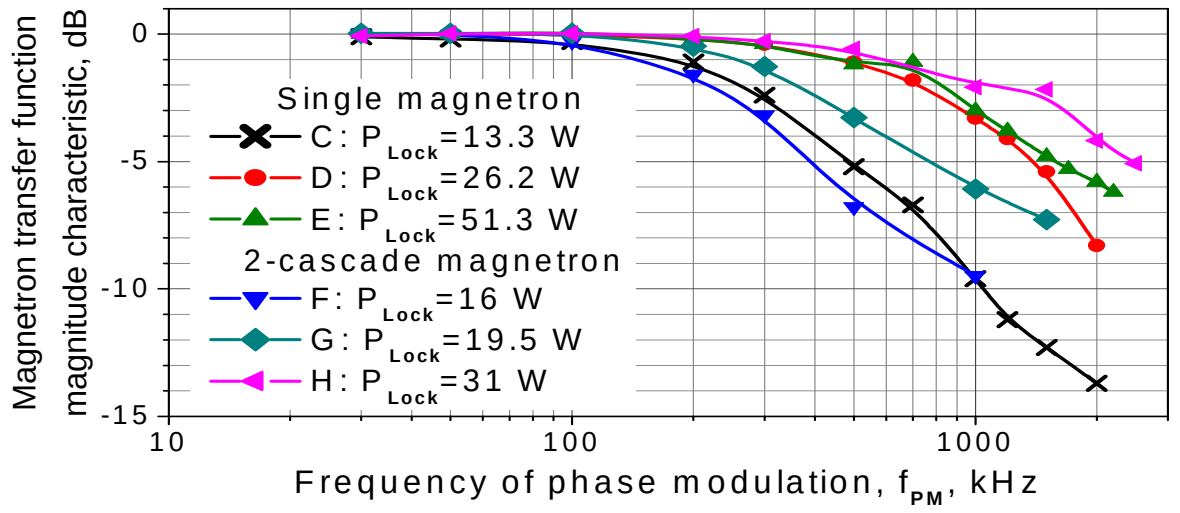

Fig. 3. Transfer function magnitude characteristics of the phase control measured in the phase modulation domain with single and 2-cascade injection-locked magnetrons at various power levels of the injection-locking signal, $P_{Lock}$.

The phase response of the magnetrons to the phase-modulated signal vs. frequency of the phase modulation, $f_{PM}$, has been measured with the calibrated phase detector including the phase shifter φ, double balanced mixer and Low Pass Filter, LPF, Fig.1, at the magnitude of the modulation of 0.35 rad., and the magnetron output power of $P_{Out}$ ≈500 W [6]. The transfer function phase characteristics (phase Bode plots) for the single and 2-cascade injection-locked magnetrons at various power values of the locking signal are shown in Fig. 4. The plots consider the magnitude response of magnetrons on the phase-modulated locking signal and the phase detector instrumental function.

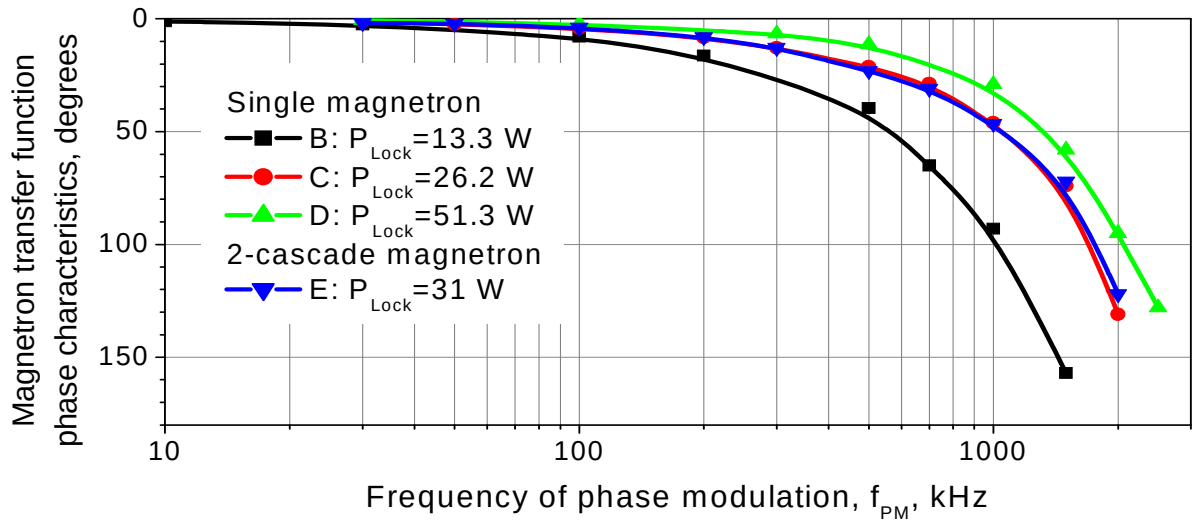

Fig. 4. Transfer function phase characteristics of single and 2-cascade magnetrons injection-locked by the phase-modulated signal vs. the modulating frequency, $f_{PM}$.

In accordance with the standard criteria (-3 dB for the magnitude characteristics and 45 degrees for the phase characteristics) the plots demonstrate an allowable bandwidth of the feedback loops control exceeding 1 MHz at $P_{Lock}$ ≈30 W for a single 1 kW 2.45 GHz magnetron. For a 2-cascade magnetron with output power of 1 kW the same bandwidth of the feedback loop will be at the injected power $P_{Lock}$ ≈1.6 W (considering the attenuator). The plots in Fig. 4 show that the phase deviation of the output magnetron signal from the phase of the injection-locking signal become noticeable beyond 10 kHz and increases with the frequency of the phase modulation or/and with a decrease of power of the locking signal.

Expressing the ratio of the magnetron (nominal) power to power of the resonant driving (injection-locking) signal as $G=P_{Nom}/P_{Lock}$ one can say that the 1 MHz bandwidth of the feedback loops for the 2.45 GHz magnetrons is provided by $G$ ~10-13 dB per cascade (tube).

The vector methods of power control in magnetrons are reduced to phase control by the phase-modulated injection-locking signal. The vector method managing the depth of the phase modulation to control the magnetron power was tested with a single-cell 2.45 GHz SRF cavity powered by the phase- and power-controlled single CW magnetron type 2M137-IL. The measured rms phase and amplitude deviations of the accelerating field in the cavity at 4K did not exceed 0.26 deg. and 0.3%, respectively, at the feedback loops bandwidth of about 100 kHz [7]. The results verify that the magnetron transmitters controlled in a wide band by a resonant injected phase-modulated signal satisfy to requirements of superconducting accelerators. Optimization of the controlling signal and a technique for increasing the average efficiency at the wide-range power control are considered below. This is substantiated by the proposed kinetic model and the experimental studies.

# Interaction of the drifting charge with the synchronous wave in magnetrons

We consider the simple analytical kinetic model based on the charge drift approximation for conventional magnetrons driven by a resonant RF signal [8]. We discuss a conventional CW magnetron with $N$-cavities ($N$ is an even number), Fig. 5, with a constant uniform magnetic field $H$, above the critical magnetic field. The magnetron operates in the $\pi$-mode, i.e. with the RF electric field shifted by $\pi$ between neighbouring cavity gaps.

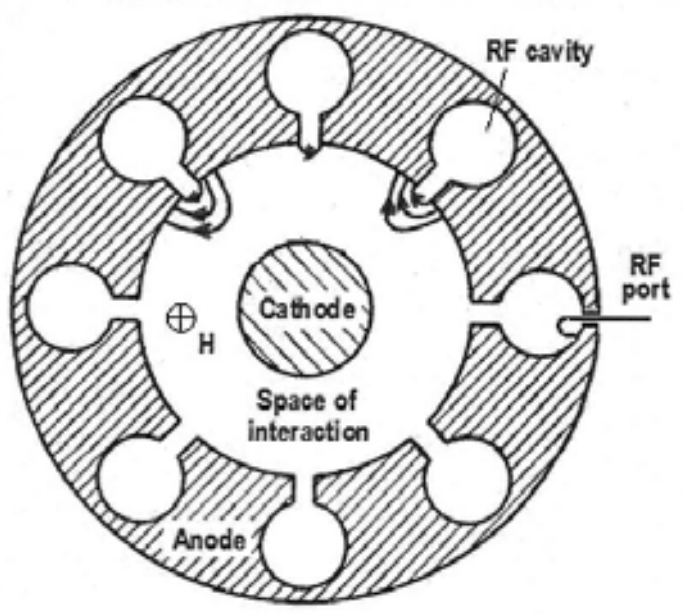

Fig. 5. Schematic sketch of an 8-cavity conventional magnetron. The lines in the space of interaction represent the RF electric field.

We consider the magnetron operating at the frequency $\omega$, being loaded by a matched load with negligible reflected signal. In the drift approximation, we consider the motion of charge in the center of the Larmor orbit with radius $r_L$, when the Larmor motion of the electron itselvf is averaged over the cyclotron frequency $\omega$ in the magnetron space of interaction, (see [9]). We neglect, as in this reference, the impact of space charge, especially since we consider the CW tubes operating at rather low currents. In this approach, neglecting the azimuthal non-uniformity of the static electric field, the drift of the center of the Larmor orbit with azimuthal angular velocity $\Omega$ in the uniform magnetic field is determined by the superposition of the static electric field described by the static electric potential $\Phi^0$ and the RF field of the synchronous wave induced by the magnetron current and the injected resonant RF signal (in a steady-state both are in phase), determined by a scalar potential $\Phi$. Thus, the drift of the charge can be described in the polar frame by the following system of equations of the first order [10]:

$$\begin{cases} \dot{r} = -\dfrac{c}{Hr}\dfrac{\partial}{\partial\varphi}\left(\Phi^0 + \Phi\right), \\ \dot{\varphi} = \dfrac{c}{Hr}\dfrac{\partial}{\partial r}\left(\Phi^0 + \Phi\right) \end{cases} \quad (1)$$

For the static electric field: $\Phi^0 = U\ln(r/r_1)/\ln(r_2/r_1)$ and $E_r$=grad$\Phi^0$, $\partial\Phi^0/\partial\varphi = 0$, therefore, $E_\varphi(r)$=0. Here $U$ is the magnetron feeding voltage; $r_1$ and $r_2$ are the magnetron cathode and anode radii, respectively.

We consider a slow RF wave type $\exp\left(-i(n\varphi + \omega t)\right)$ excited at the frequency $\omega$ and rotating in the space of interaction with the phase velocity $\Omega=\omega/n$, [9]. The wave number $n=N/2$ sets the same azimuthal periodicity in interaction of the drifting charge with the RF field in the space of interaction. The phase velocity of the wave coincides with the azimuthal drift velocity of the center of the Larmor orbit located on the "synchronous" radius, $r_S$.

In magnetrons $r_L << 2\pi c/n\Omega$ (the right part of the inequality is the length of the synchronous wave). Therefore one can consider the interaction of the synchronous wave with an electron rotating along the Larmor orbit as an interaction of the wave with the point charge located in the center of the orbit. The electric field of the synchronous wave has radial and azimuthal components whose sign and magnitude are determined by the phase of the wave.

In a conventional magnetron ($\omega << \pi c/r_1$, $\pi c/r_2$), thus the quasi-static approximation can be used to describe the rotating synchronous wave in the magnetron space of interaction. The scalar potential $\Phi$, satisfying the Laplace equation for the rotating wave is presented as in ref. [8]:

$$\Phi = \sum_{k=-\infty}^{\infty} \frac{\widetilde{E}_k \cdot r_1}{2k}\left[\left(\frac{r}{r_1}\right)^k - \left(\frac{r_1}{r}\right)^k\right]\sin(k\varphi + \omega t) \quad , \quad (2)$$

where $\tilde{E}_k$ is the amplitude of the $k$-th harmonic of radial RF electric field at $r = r_1$. The form of the potential was chosen so that the azimuthal electric field vanishes at the cathode. The coefficients $\tilde{E}_k$ are determined by the requirement to have zero azimuthal electric field at the anode everywhere except the coupling slits of the cavities. The term in the sum of Eq. (2) with $k=n$ has a resonant interaction with the azimuthal motion of the Larmor orbit. As such, we consider only this term. Note that from Eqs. (1) it follows that without the synchronous wave $\dot{r} = 0$ and a drift of the charge towards the anode is impossible.

In the coordinate frame rotating with the synchronous wave for $\varphi_S=\varphi + t\cdot\omega/n$ and for an effective potential $\Phi_S$:

$$\Phi_S = U\frac{\ln(r/r_1)}{\ln(r_2/r_1)} + \frac{\omega H}{2nc}r^2 + \frac{\widetilde{E}_n r_1}{2n}\left[\left(\frac{r}{r_1}\right)^n - \left(\frac{r_1}{r}\right)^n\right]\sin(n\varphi_S)$$

one obtains the system of drift equations, [10]:

$$\begin{cases} \dot{r} = -\dfrac{c}{Hr}\dfrac{\partial}{\partial\varphi_S}\Phi_S \;, \\ \dot{\varphi}_S = \dfrac{c}{Hr}\dfrac{\partial}{\partial r}\Phi_S \end{cases} \quad (3)$$

Substituting the potential $\Phi_S$ into Eqs. (3) and denoting

$$\phi_0(r) = \ln\frac{r}{r_1} - \frac{1}{2}\left(\frac{r}{r_S}\right)^2, \quad \phi_1(r) = \frac{1}{2n}\left[\left(\frac{r}{r_1}\right)^n - \left(\frac{r_1}{r}\right)^n\right],$$

one can obtain the system of drift equations, [10, 8] in the frame of the synchronous wave expressed via the relative magnitude of the resonant harmonic of the synchronous wave, $\varepsilon$, taken in the neighbourhood of the cathode:

$$\begin{cases} \dot{r} = \omega\dfrac{r_S^{\,2}}{r}\varepsilon\phi_1(r)\cos(n\varphi_S) \\ n\dot{\varphi}_S = -\omega\dfrac{r_S^{\,2}}{r}\left(\dfrac{d\phi_0}{dr} + \varepsilon\dfrac{d\phi_1}{dr}\sin(n\varphi_S)\right) \end{cases} . \quad (4)$$

Here $r_S = \sqrt{-ncU/\left(\omega H\ln(r_2/r_1)\right)}$ is the "synchronous" radius ($H$<0 is assumed), $\varepsilon = \widetilde{E}_n/E_1 = \widetilde{E}_n \cdot r_1\ln(r_2/r_1)/U$.

The simplified Eqs. (4) do not describe the coherent oscillation in a magnetron, but they characterize the resonant interaction of the charge in the center of the Larmor orbit with the synchronous wave. This is the basis of magnetrons operation.

The top equation in Eqs. (4) describes the radial velocity of the moving charge. In accordance with this equation, the drift of the charge towards the anode is possible at $-\pi/2 < n\varphi_S < \pi/2$ with a period of 2π, *i.e.*, only in "spokes". The charge can enter the "spoke" through the boundaries located at $\pm\pi/2$, [10]. The radial drift velocity is proportional to the synchronous wave magnitude, $\varepsilon$. The condition $\varepsilon \geq 1$ does not allow operation of the magnetron.

The second equation describes the azimuthal velocity of the drifting charge in the frame of the synchronous wave. The second term in the parentheses causes phase grouping of the charge by the resonant RF field via the potential $\phi_1$.

The first term of the equation describes a radially-dependent azimuthal drift of the charge resulting from the rotating frame with azimuthal angular velocity $-\omega/n$. This term at $r > r_S$ and at low $\varepsilon$ causes the movement of the charge from the phase interval ±π/2 allowed for "spokes", [8].

The Eqs. (4) were integrated for a typical model of a commercial magnetron described in ref. [8] with $N$=8, $r_1$ = 5 mm, $r_2/r_1$ =1.5, $r_S/r_1$ =1.2. Considering the charge drifting in the center of Larmor orbit we integrated the charge trajectories at $r \geq r_1+r_L$ for various magnitudes $\varepsilon$ of the RF field in the synchronous wave and at the time interval ($\tau$) of the drift during 2-10 cyclotron periods allowing coherent contribution to the synchronous wave, [9]. The azimuthal boundaries of the charge drifting in a "spoke" at various $\varepsilon$ obtained by the integration are plotted in Fig. 6A. The phase interval at $(r_1+r_L)$ normalized by $\pi$ indicates part of charge reaching the anode at the considered $\tau$ value. The graphs show large losses and insufficient grouping of the charge contributing to the coherent radiation at $\varepsilon \leq 0.2$,, while an increase of $\varepsilon$ to the value of $\varepsilon \cong 0.3$ (*e.g.* due to the injected resonant signal) decreases loss of charge [8], and improves the phase grouping. Fig. 6B shows trajectories of the charge in a "spoke" at $\varepsilon$=0.3.

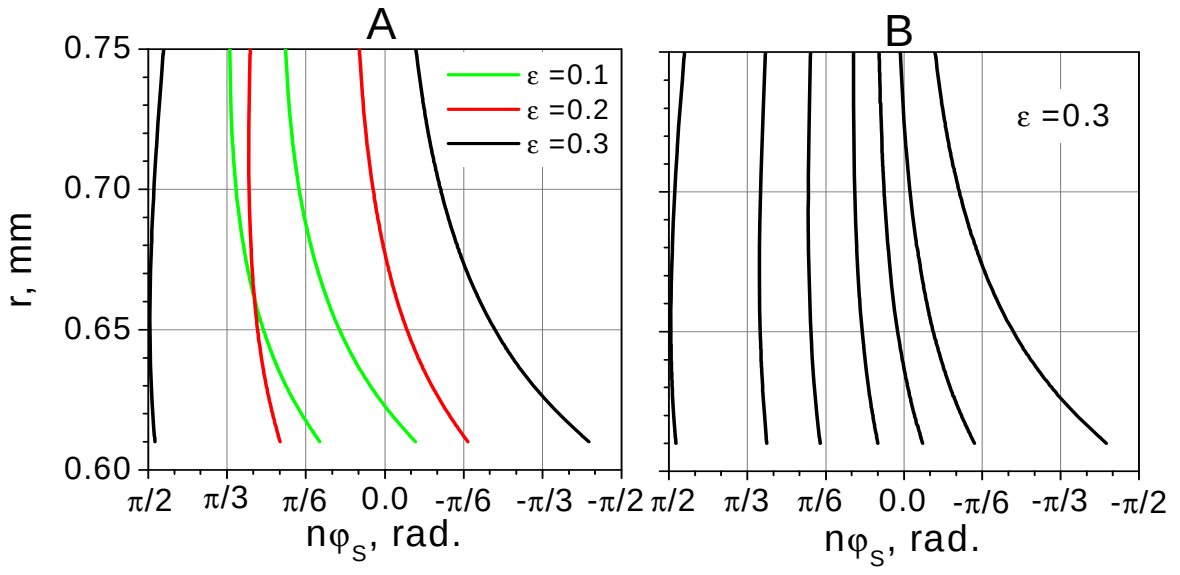

Fig. 6. Grouping of the charge drifting towards the magnetron anode in the considered magnetron model.

The resonant phase grouping in magnetrons can be explained in the following manner. In the frame of the slow synchronous wave the RF azimuthal electric field in a "spoke" can be considered as stationary. The electric field strongly coupled with the resonant mode of the magnetron oscillation acts on the charge drifting in the interaction space. This causes the resonant energy exchange between the synchronous wave and the charge. If the azimuthal velocity of the drifting charge is greater than the azimuthal velocity of the synchronous wave, the charge induces oscillation of the resonant mode in the magnetron RF system, being decelerated, [11], and contributes it to the synchronous wave. This increases the synchronous wave amplitude. Otherwise, the electric field of the wave accelerates the charge increasing its azimuthal drift velocity. This reduces the wave energy and the amplitude, respectively. In other words, the resonant phase grouping in magnetrons is realized by variation of the azimuthal drift velocity of the charge due to variation of the radial component of the electric field of the synchronous wave caused by the energy exchange between the wave and the charge drifting in crossed stationary fields, [12]. Thus, the increase or decrease of the self-consistent electric field of the synchronous wave in a magnetron, resulting from the continuous resonant energy exchange, is determined by the difference in the azimuthal velocities of the drifting charge and the synchronous wave, $\Delta v_{AZ}/c$, along the trajectories of the charge drift in a "spoke", Fig. 7. As a result of the phase grouping, the graphs indicate the non-isotropic distribution of the azimuthal charge drift velocity. We note once again that the energy exchange between the synchronous wave and the averaged motion of Larmor electrons in magnetrons is considered in the drift approximation as the energy exchange between the wave and charges drifting in the centers of Larmor orbits.

Note that the azimuthal component of the synchronous wave provides the radial drift of the charge.

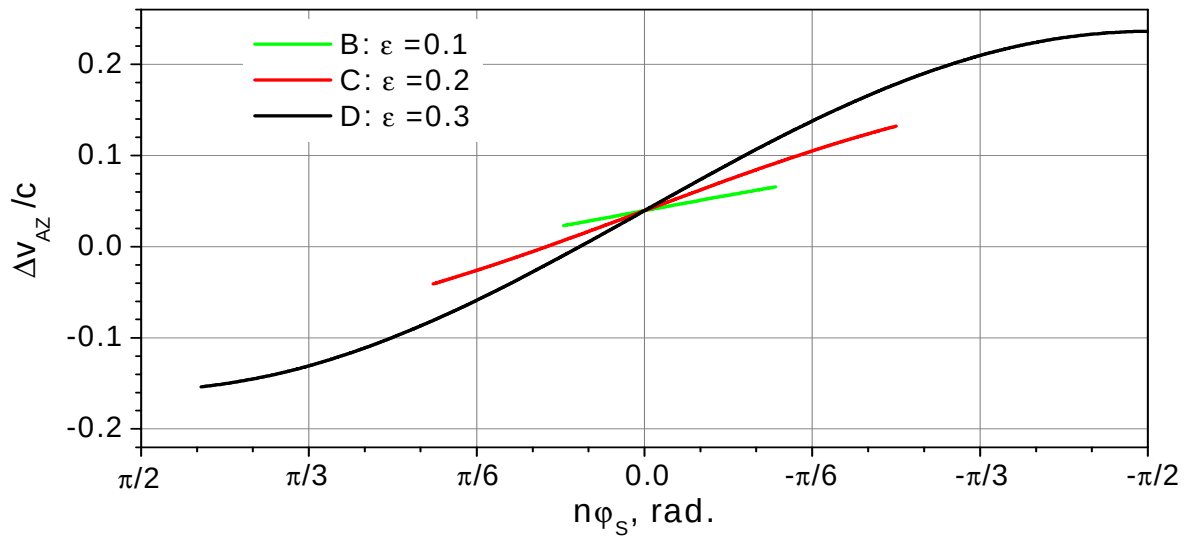

Fig. 7. Distributions of the differences of the azimuthal velocities of the drifting charge and the synchronous wave, $\Delta v_{AZ}/c$, at various $\varepsilon$. The solid lines represent the phase intervals (at various $\varepsilon$), in which the drifting charge contributes to the coherent radiation.

The loss of charge drifting towards the anode, Figs. 6A and 7, reduces the magnetron current in "spokes", *i.e.* the magnitude of the induced synchronous wave. The charge lost at low $\varepsilon$ migrates from a "spoke" into the adjacent phase interval [8], where it is accelerated by the synchronous wave. This also reduces the synchronous wave magnitude, deteriorating phase grouping. Thus, for stable operation of the magnetron, one needs to prevent a decrease of $\varepsilon$, otherwise the insufficient magnitude of the synchronous wave causes noisy operation of the tube and, in principle, may lead to disruption of coherent generation resulting from loss of coherency as it is shown below.

Graphs in Fig. 7 indicate that the increment in the synchronous wave energy at $\varepsilon \leq 0.2$ cannot compensate the wave energy decrement. This results in non-functioning of the tube as a coherent oscillator, while at $\varepsilon \cong 0.3$ the compensation allows operation of the considered magnetron model.

Moreover, the acceleration of the loosed charge increases the Larmor radius of the rotating electron which may hit the cathode increasing the magnetron cathode losses. For the typical magnetron model, [8], the losses of the drifting charge are minimized at $\varepsilon$~0.3 in the main part of the charge drift trajectories. Thus, a sufficient magnitude of the synchronous wave protects from loss of coherency and prevents an increase of the cathode losses even at a low magnetron power.

A resonant driving signal injected in the magnetron in accordance with the energy conservation law increases the RF energy stored in the magnetron cavities and in the interaction space. Since the RF energy in the magnetron is determined by the static electric field, the injected resonant signal is equivalent to an increase of the magnetron feeding voltage. Thus, a sufficient power of injected resonant signal allows the magnetron to start-up even if the magnetron feeding voltage is somewhat less than the threshold of self-excitation. In this case the magnetron current is less than the minimum current when the tube is self excited (the free run operation) [8]. This allows stable operation of the magnetron at an extended range of current (power) control. A lack of RF voltage in the synchronous wave induced by the lower magnetron current is compensated by the injected resonant signal providing stable operation of the tube. As follows from the charge drift model estimations and experiments with the 2.45 GHz, 1 kW magnetrons, the injected resonant signal with a power of about -10 dB of the nominal magnetron power allows magnetron power control over the range of 10 dB by deep variation of the magnetron current [8]. Operation of the magnetron at a current less than the allowable minimum current for free run operation in accordance with its the Volt-Amp (V-I) characteristic is possible at a magnetron voltage below the threshold of self-excitation. At a sufficient power of the resonant driving signal the loss of the drifting charge and the cathode losses remain low even at a small magnetron current. Thus, one can expect the highest efficiency of a magnetron driven by a sufficient resonant signal providing stable operation of the tube in an extended range of current (power) control.

## Impact of magnitude of the injection-locking signal on the magnetron operation

How the magnitude of the injected resonant signal affects the magnetron operation was studied with a 2.45 GHz, 1.2 kW magnetron type 2M137-IL with a permanent magnet operating in the CW regime, [8], as it is shown in Fig. 8.

In accordance with the setup the magnetron was started up and frequency-locked by the HP 8341A generator via a Solid-State Amplifier (SSA) and 36.6 dB TWT amplifier providing CW locking power up to 100 W. The magnetron was powered by an Alter switching High Voltage (HV) power supply type SM445G with a current feedback loop, operating as a current source and allowing current control. The setup was used to measure the magnetron power, efficiency, the noise at various levels of powers of the magnetron and the locking signal and stability of the carrier frequency of the tube.

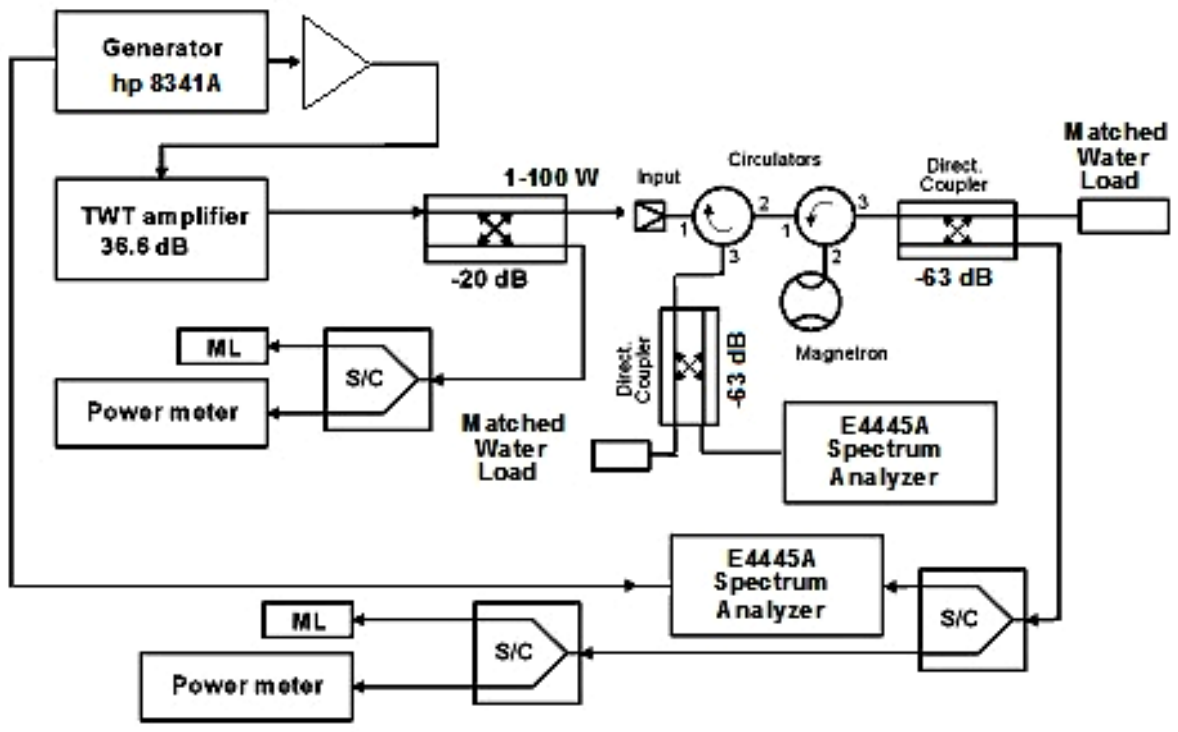

Fig. 8. Schematic of the magnetron CW setup to test the magnetron efficiency, noise and frequency stability, [8].

The V-I characteristic of the magnetron was measured in the CW regime. The magnetron cathode high voltage and current have been measured by the calibrated compensating divider and the transducer, respectively, via a scope. The inaccuracy of the calibration did not exceed ±1%. Start up and injection-locking of the magnetron at a current less than the minimum current in free run operation was performed by injection of the resonant signal with power $P_{Lock}$ =100 W. Fig. 9 shows the V-I characteristic with standard deviation error bars.

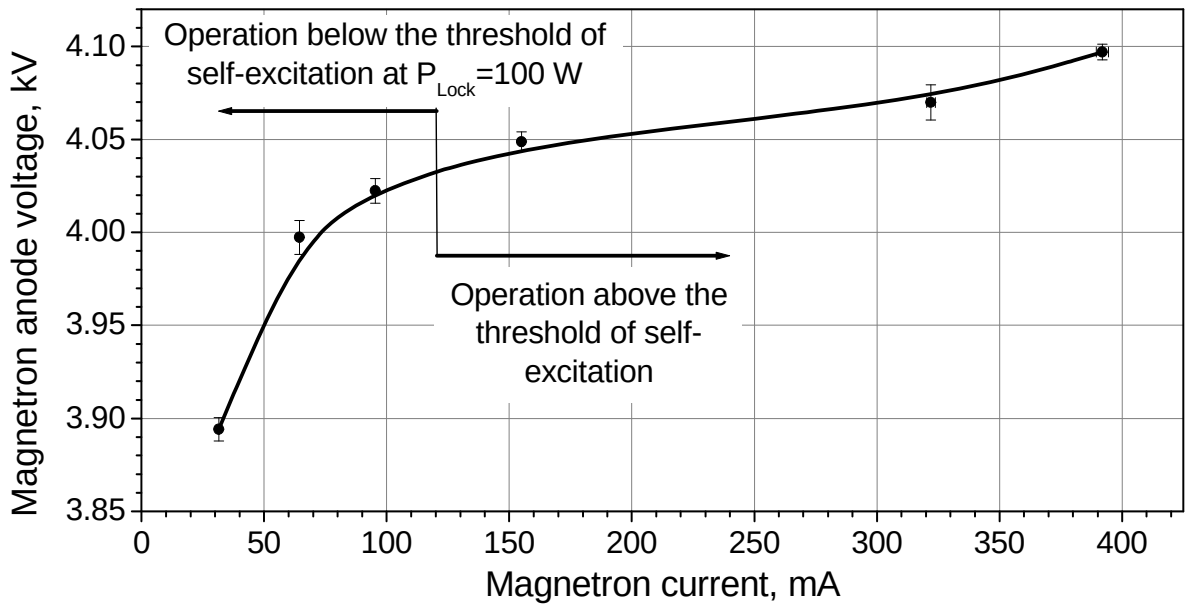

Fig. 9. The magnetron V-I characteristic measured at $P_{Lock}$ =100 W. The solid line (B-spline fit) shows the available range of current with stable operation of the tube at the given $P_{Lock}$=100 W.

As it was discussed above, a power level of the injected resonant signal of about -10 dB of the magnetron nominal power is sufficient to start up the tube providing stable operation at currents over the range of 31.6-392 mA, *i.e.*, in the range of power control ≥10 dB.

The power of the coherent synchronous wave is proportional to the squared charge reaching the magnetron anode. Therefore, a reduction of the drifting charge at a small injection-locking signal greatly increases the relative power fluctuations in the synchronous wave and in the entire magnetron RF system increasing the magnetron noise. A significant loss of the drifting charge may lead to loss of coherency in the generation of the tube, turning it into noise because of damping of the synchronous wave. A similar phenomenon is observed in the work of other coherent generators with a phase grouping of the beam by a synchronous wave, *e.g.*, in free electron lasers, etc.

We studied this phenomenon in two sets of experiments measuring the spectral density of the power of the magnetron noise relative to the carrier frequency power.

In the first set, we studied the operation of the magnetron below the threshold of self-excitation at an output power of 100 W, Fig. 10, when $\varepsilon$ is generally determined by the injected resonant signal and one can assume that the fluctuations of $\varepsilon$ are quite small due to the stability of the injected resonant signal.

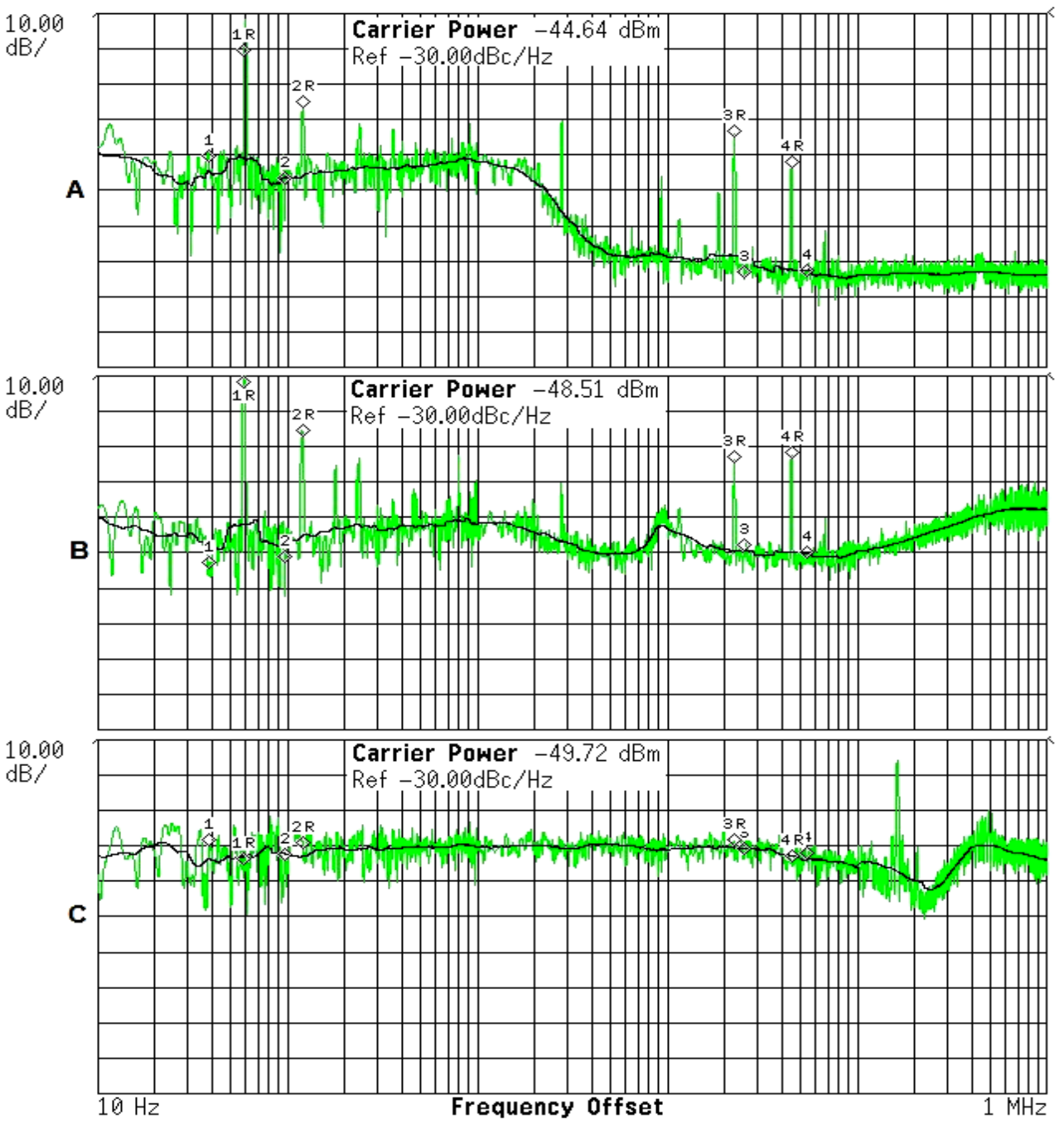

Fig. 10: The spectral power density of the noise at various power levels of the locking signal at $P_{Mag}$ = 100 W. A - $P_{Lock}$ =100 W, B - $P_{Lock}$ =30 W, C - $P_{Lock}$ =10 W, Black traces show the averaged spectral power density of the noise.

The magnetron being started up and injection-locked at various powers of the locking signal shows a dramatic increase of measured density of the noise power in the range from 1 kHz to 1 MHz if the injected resonant power is decreased from 100 W to 10 W, Fig. 10, traces A, B, C.

The dramatic increase of the noise power density beginning at 1 kHz or less (that is much less then the switching frequency of the HV power supply) can be explained as relaxation oscillations with a characteristic frequency of $f_c \sim 1/(Z_S \cdot C_c)$ resulting from partial or total loss of coherency in magnetron oscillations at insufficient $\varepsilon$ values, traces B and C, respectively. Here $Z_S$ is the magnetron static impedance and $C_c$ is the capacitance of the magnetron cathode HV circuitry. The relaxation oscillations are caused most likely by a short overvoltage in the magnetron resulted from a disruption of the coherent oscillation; this may restore conditions for a start up of the tube with an appropriate phase grouping during a short time. For the magnetron operating with power of 100 W, $Z_S$ is about 100 kOhms, Fig. 9. This corresponds to $f_c \geq 1$ kHz at $C_c \leq 10$ nF.

Note that the sidebands shown in the magnetron spectra in Figs. 10 and 11 result from the magnetron switching power supply and switching power supply of the TWT amplifier. The sidebands caused by the TWT switching power supply are seen in Fig. 11, trace D, showing the spectral density of the noise power of the injection-locking signal when the magnetron high voltage was OFF.

Measured spectral power densities of the noise vs. power of the locking signal at the nominal magnetron power, when the injected resonant signal less affects the phase grouping, are plotted in Fig. 11.

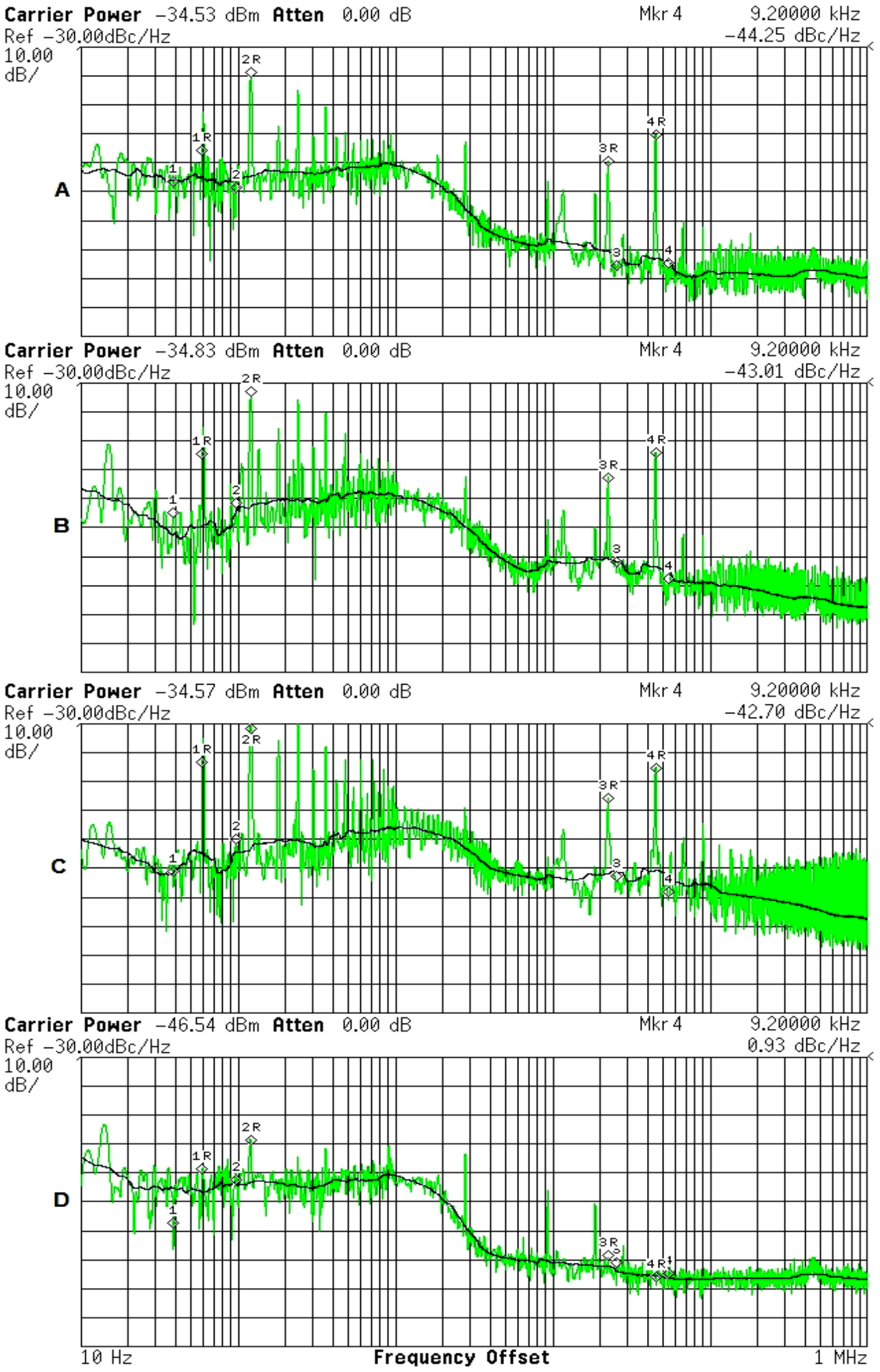

Fig. 11. The spectral power density of the noise of magnetron at the output power of 1 kW, at the power of the locking signal of 100, 30 and 10 W, traces A, B and C, respectively, [8]. Traces D are the spectral power density of noise of the injection-locking signal ($P_{Lock}$=100 W), when the magnetron feeding voltage is OFF. Black traces show the averaged spectral power density of the noise.

At a low locking power the fluctuations of $\varepsilon$ are approximately proportional to fluctuations of the charge reaching the anode; they cause amplitude modulations of the synchronous wave with frequencies $f_M$ integrated by the magnetron cavity with Q-factor $Q_M$ over the time $\tau_M \sim Q_M /(\pi \cdot f_M)$. Thus the noise caused by amplitude fluctuations in synchronous wave is limited by MHz range.

The traces A-C in Fig. 11 illustrate a notable reduction of the magnetron spectral power density of the noise (by ~20 dBc/Hz) at frequencies > 100 kHz when the power of the injection-locking signal is increased from 10 W to 100 W. This indicates a reduction in fluctuations of $\varepsilon$, by an increase of contribution of the stable injection-locking signal.

Comparison of traces B and C in Fig. 10 with traces B and C in Fig. 11, measured at the same power of the locking signal ($P_{Lock}$=30 W and 10 W, respectively) demonstrates a significant difference of the spectral power density in the low-frequency range (< 100 kHz) for the low output power (100 W) and the high output power (1 kW) of the magnetron.

In the frequency range of (1-100 kHz) , Fig. 11, traces A-C, the decrease of the locking power from -10 dB to -20 dB increases the power density of noise by ~20 dBc/Hz. This indicates deterioration (partial losses) of coherency at the almost nominal power of the tube driven by the insufficient injection-locking signal. Disruption of coherent generation of the magnetron caused by total loss of coherency, as it was observed at low powers of the tube and the locking signal (Fig. 10 C) at the almost nominal magnetron power and the low power of locking signal was not observed.

Thus, the noticeable fluctuations of the charge reaching the magnetron anode cause fluctuations in the magnitude of the synchronous wave resulting in noise in the MHz range. The noise is effectively suppressed by increasing the injection-locking signal to -10 dB.

At a magnetron voltage somewhat less than the threshold of self-excitation, the tube power is low and the magnetron demonstrates loss of coherency at an insufficient injected resonant signal. This is manifested in a dramatic increase of noise in a low-frequency range resulting from a significant deterioration of the phase grouping, traces B, and a total disruption of coherent oscillation, traces C, respectively in Fig. 10.

At the injected resonant signal of -10 dB the magnetron demonstrates low noise (less than -50 dBc) avoiding loss of coherency for the output power ranging from 100 W to 1000 W, Figs. 10 and 11, traces A.

The magnetron absolute efficiency, η, was determined as the ratio of the measured magnetron RF power to the output power of the magnetron HV power supply at various values of the magnetron RF power [8]. Fig. 12 shows the magnetron average efficiency in dependence on the range of power control (in dB) assuming a linear variation of the magnetron power in the given range. The graphs demonstrate highest average efficiency of the magnetron injection-locked by a sufficient driving signal at a wide range of (current) power control in comparison with the vector power control methods.

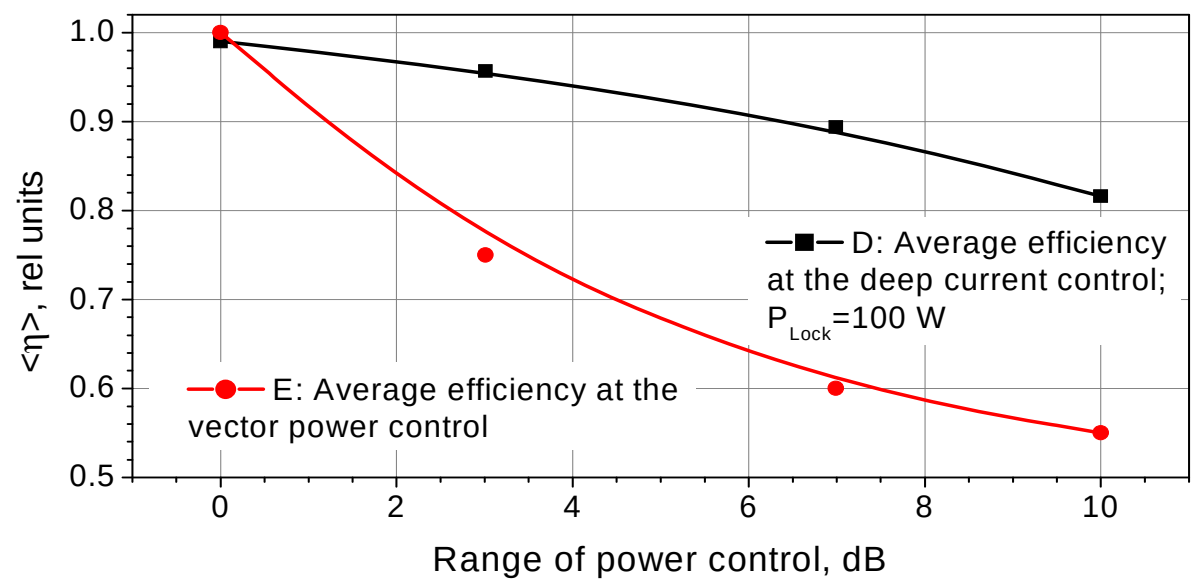

Fig. 12. Relative averaged magnetron efficiency vs. range of power control for various methods of control. Curve D shows the average efficiency of the 1.2 kW magnetron driven by the injected resonant signal of -10 dB and measured at deep magnetron current control. Curve E shows the average efficiency of 1 kW magnetrons with vector power control [6, 7].

Precise stability of the carrier frequency at operation of the magnetron in the regime with a wide-range power control at a sufficient locking signal is demonstrated in Fig. 13, [8], which shows the offset of the carrier frequency at various power levels of the magnetron and the locking signal.

The measured offset of the carrier frequency does not demonstrate any broadening of the magnetron spectral line over a wide range of the magnetron power control. This implies that no noticeable noise of the magnetron is produced when operating below the threshold of self-excitation at a sufficient driving resonant signal and no losses of coherency are encountered.

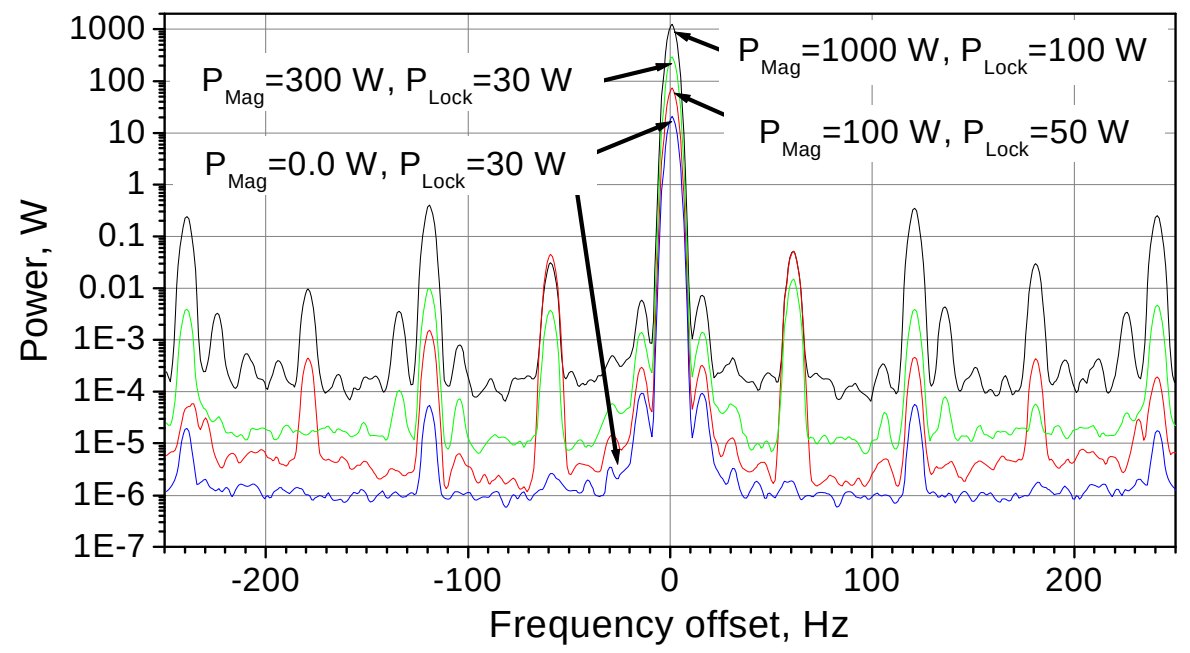

Fig. 13. Offset of the carrier frequency at various power levels of the magnetron, $P_{Mag}$, and the locking signal, $P_{Lock}$. The trace $P_{Mag}$ =0.0 W, $P_{Lock}$ =30 shows the frequency offset of the injection-locking signal.

Shown in the plots, trace $P_{Mag}$ =0.0 W, $P_{Lock}$ =30 W was measured with the magnetron high voltage turned OFF.

The low frequency sidebands in all traces are caused by 60 Hz modulation of the switching power supplies of the magnetron and TWT.

## Modelling of a dynamic power control by management of magnetron current

The capability of the proposed method for a deep dynamic power control was verified by a modulation of the magnetron current via control of the HV switching power supply within a current feedback loop, Fig. 14, [13].

The low-frequency harmonic analogue signal driving the control input of the SM445G power supply regulated the current of the magnetron driven by the injected resonant RF signal at $P_{Lock}$ =100 W. The magnetron power was determined by measurements of the magnetron current with a calibrated transducer and by measurements of RF power vs. the magnetron current. The inaccuracy of the calibration and the RF power measurements did not exceed ±1%. The traces were averaged over 16 runs reducing the noise caused by operation of the switching power supply.

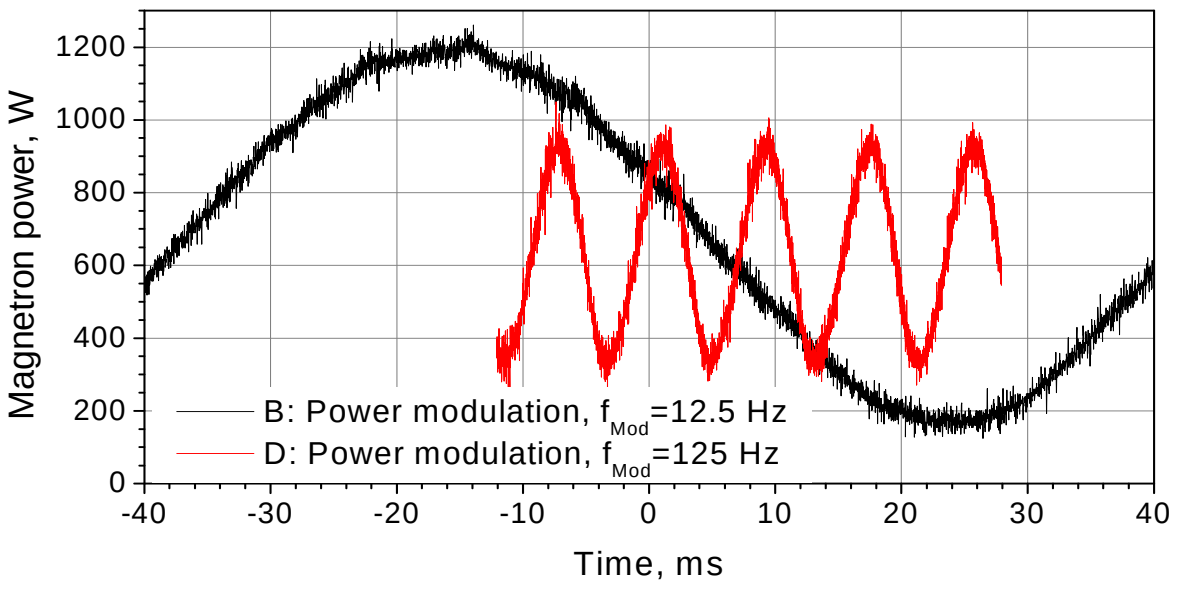

Fig. 14. Modulation of the magnetron power managing the magnetron current by a harmonic signal controlling the SM445G HV switching power supply.

The bandwidth of the power control was limited by the bandwidth of the power supply feedback loop since the power supply was designed for slow current control. Nevertheless, the harmonic shapes of the measured traces are a proof-of-principle of wide range dynamic power control by a wide-range management of the magnetron current.

## Conceptual scheme of highly-efficient high-power magnetron transmitter

A conceptual scheme of a single-channel transmitter based on magnetrons, allowing dynamic wideband phase and mid-frequency wide-range power management at the highest average efficiency is shown in Fig. 15, [13].

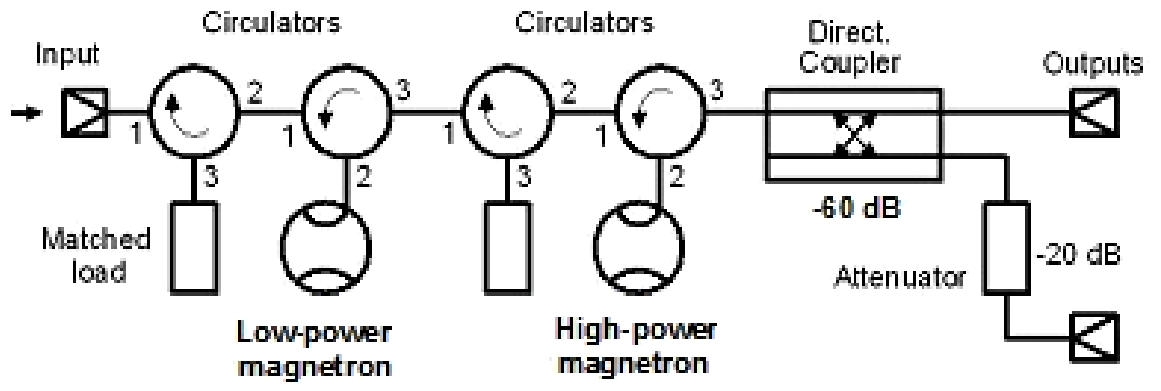

Fig. 15. Conceptual scheme of a 2-cascade magnetron single-channel transmitter allowing dynamic phase and power control at highest efficiency.

In this scheme the first, low-power magnetron, provides phase modulation (control) of the signal frequency-locking the second, high-power magnetron. Power control in the required range (up to 10 dB) is realized by modulation (control) of current in the high-power magnetron which at low current operates at a voltage less than the critical voltage in free run mode. Stable and low noise operation of the high-power injection-locked magnetron at a voltage less than the critical in free run is provided by a resonant injected (injection-locking) signal with sufficient power.

The current management may be provided by a high-voltage switching power supply within a current feedback loop. Presently the bandwidth of such a magnetron power (current) control can be up to 10 kHz without compromising the power supply efficiency at it was noted above. This bandwidth is sufficient for various SRF accelerators with beam currents ≥5 mA, *e.g.*, for ADS-class projects. Thus, a single-channel magnetron transmitter with a sufficient injection-locking signal (about -10 dB), as substantiated by the presented experimental results, will allow dynamic wideband phase and mid-frequency power control in ADS-class projects.

The control (modulation) of the magnetron current causes phase pushing in the frequency-locked magnetron. At a bandwidth of the phase control in the MHz range, one expects the phase pushing to be eliminated to a level less than -50 dB, suitable for various SRF accelerators.

As it was shown in Figs. 2 and 3, the 2-cascade transmitters based on the injection-locked magnetrons provide bandwidth of phase control in the MHz range at a G value of about 20-25 dB, *i.e.*, 10-12 dB per cascade. As was demonstrated above and in ref. [8], this value is also sufficient for operation of a single-channel 2-cascade magnetron transmitter with power control in the range up to 10 dB by control of the magnetron current over an extended range.

Thus a transmitter with the conceptual scheme presented above, satisfying requirements for various superconducting accelerators, will provide the highest efficiency at the lowest capital and operation costs adequate for the requirements of various superconducting accelerators. The highest efficiency is very important for ADS class projects. The developed technique of the magnetron power control by the deep variation of the magnetron current can be combined with the vector methods of power control. In this case the transmitter will provide wideband phase and power control at the highest efficiency.

## Summary

The presented simple kinetic model and analysis of the drift approximation clarifies the physics of operation of the injection-locked magnetrons; modelling the interaction of the drifting charge with the slow synchronous wave excited in the tubes that is the basis of operation of magnetrons. Based on the modelling, estimations, described in [8], and the presented analysis, we proposed and tested impact of increasing to -10 dB the injected resonant signal driving the magnetron. The experimental results are in agreement with the modelling and demonstrate the capability of the proposed magnetron transmitter for the dynamic phase and power control required for superconducting accelerators. The impact of the power of the injection-locking signal on the magnetron noise, average efficiency at a wide range of power control, and stability in operation in agreement with the drift approximation model was verified in the described experiments. Utilization of the obtained results allows for optimization of the magnetron transmitter parameters in choosing the conceptual design that is optimal for superconducting accelerators. A significant improvement of the average efficiency of the magnetron transmitter with dynamic phase and power control will allow a significant decrease in the capital and operation costs of large-scale high-current superconducting accelerator projects including ADS projects. The proposed conceptual scheme of the highly-efficient and cost-effective magnetron transmitter based on the obtained results is adequate for the requirements of various superconducting accelerators.

# Acknowledgement

We are very thankful to Dr. Helen Edwards for permanent interest and support of this work, Mr. Brian Chase and Mr. Ralph Pasquinelli for help in experiments and Dr. Ya. Derbenev, Dr. R. Thurman-Keup, Dr. F. Marhauser and Dr. Yu. Eidelman for fruitful discussions.